\documentstyle[12pt,epsfig]{article}
\newcommand{\nd}{\noindent}
\newcommand{\beq}{\begin{equation}}
\newcommand{\eeq}{\end{equation}}
\newcommand{\barr}{\begin{eqnarray}}
\newcommand{\earr}{\end{eqnarray}}
\newcommand{\ba}{\begin{array}}
\newcommand{\ea}{\end{array}}

\newcommand{\bfp}{\mbox{\boldmath $p$}}
\newcommand{\bfq}{\mbox{\boldmath $q$}}
\newcommand{\bfP}{\mbox{\boldmath $P$}}
\newcommand{\bfk}{\mbox{\boldmath $k$}}
\newcommand{\bfx}{\mbox{\boldmath $x$}}
\newcommand{\bfy}{\mbox{\boldmath $y$}}
\newcommand{\bfz}{\mbox{\boldmath $z$}}

\newcommand{\bfel}{\mbox{\boldmath $\ell$}}

\newcommand{\bfAN}{\mbox{\boldmath $A_N$}}

\newcommand{\pup}{p^\uparrow}

\newcommand{\qup}{q^\uparrow}

\newcommand{\qdown}{q^\downarrow}
\newcommand{\NP}[1]{{\it Nucl.\ Phys.}\ {\bf #1}}
\newcommand{\ZP}[1]{{\it Z.\ Phys.}\ {\bf #1}}
\newcommand{\PL}[1]{{\it Phys.\ Lett.}\ {\bf #1}}
\newcommand{\PR}[1]{{\it Phys.\ Rev.}\ {\bf #1}}
\newcommand{\PRL}[1]{{\it Phys.\ Rev.\ Lett.}\ {\bf #1}}

\def\lsim{\mathrel{\rlap{\lower4pt\hbox{\hskip1pt$\sim$}}\raise1pt\hbox{$<$}}}
\def\gsim{\mathrel{\rlap{\lower4pt\hbox{\hskip1pt$\sim$}}\raise1pt\hbox{$>$}}}
\def\nostrocostruttino#1\over#2{\mathrel{\mathop{\kern 0pt \rlap
{\hbox{$#1$}}} \hbox{\kern-.135em $#2$}}}

\begin{document}
%%%%%%%%%%%%%%%%%%%%%%%%%%%%%%%%%%%%%%%%%%%%%%%%%%%%%%%%%%%%%%%%%%%%%%%%%%%%%%
\begin{flushright}
DFTT 06/2000 \\
%VUTH 99-yy \\
INFNCA-TH0002 \\
hep-ph/0002120 \\
\end{flushright}
\vskip 1.5cm
\begin{center}
{\bf Spin effects in the fragmentation of a transversely polarized quark}\\
\vskip 0.8cm
{\sf M.\ Anselmino$^1$ and F.\ Murgia$^2$}
\vskip 0.5cm
{$^1$ Dipartimento di Fisica Teorica, Universit\`a di Torino and \\
      INFN, Sezione di Torino, Via P. Giuria 1, I-10125 Torino, Italy\\
\vskip 0.5cm
$^2$  Istituto Nazionale di Fisica Nucleare, Sezione di Cagliari\\
      and Dipartimento di Fisica, Universit\`a di Cagliari\\
      C.P. 170, I-09042 Monserrato (CA), Italy} \\
\end{center}
\vskip 1.5cm
\noindent
{\bf Abstract:} \\ 
An azimuthal dependence of pions produced in polarized Deep Inelastic 
Scattering, $\gamma^* \pup \to \pi X$, has been recently observed 
and might be related to the so-called Collins effect. We discuss
in details, for a general spin configuration of the nucleon, 
the kinematics of the process and methods of extracting information on
the fragmentation properties of a polarized quark. Assuming that the
observed azimuthal dependence is indeed due to Collins effect, we 
derive a lower bound estimate for the size of the quark analysing power, 
which turns out to be large. 
%%%%%%%%%%%%%%%%%%%%%%%%%%%%%%%%%%%%%%%%%%%%%%%%%%%%%%%%%%%%%%%%%%%%%%%%%%%%%%%
\newpage
%%%%%%%%%%%%%%%%%%%%%%%%%%%%%%%%%%%%%%%%%%%%%%%%%%%%%%%%%%%%%%%%%%%%%%%%%%%%%%%
\pagestyle{plain}
\setcounter{page}{1}
\nd
{\bf 1. Introduction}
\vskip 6pt

Semi-inclusive polarized Deep Inelastic Scattering combines in  
a unique way spin properties of quarks inside hadrons: a leading hadron 
inside a current jet should result from a quark inside the polarized
nucleon which -- after the well known QED interaction with the
lepton -- fragments into the hadron. The number of, say, pions produced
in such a process depends both on the quark polarized distribution
and fragmentation functions. If one considers transverse polarizations,
new, yet unexplored properties become accessible, which have been or might 
soon be observed in several experiments in progress. 

An example of this is the azimuthal dependence -- with respect to the
$\gamma^*$ direction -- of pions produced in $\ell p \to \ell' \pi X$ 
processes, with unpolarized leptons and polarized protons, 
recently observed by HERMES \cite{her} and
by SMC \cite {smc} collaborations; if confirmed, this asymmetry
would prove a novel feature of the fragmentation process of a transversely 
polarized quark, namely its dependence on the direction of the intrinsic 
transverse momentum of the hadron inside the jet. Such spin and $\bfk_\perp$ 
dependence was predicted some time ago \cite{col}, but only recently became 
object of a careful investigation. 
It might be the crucial key towards a complete understanding of other 
single spin asymmetries observed in $p \pup \to \pi X$ processes \cite{e704}, 
and related to quark transverse motion \cite{siv}-\cite{noi2}. 

We concentrate here on such a single spin effect in DIS, giving explicit
expressions for the measurable observables, considering different initial 
spin states of the nucleon, discussing the implications of the existing data 
and extracting from them some phenomenological information. We do not attempt 
to explain the data using particular models for the unknown distribution or 
fragmentation functions, as other authors do \cite{kot}-\cite{oga}.
Our work is somewhat related to that of Ref. \cite{bm}, with the 
difference that we focus our attention on direct cross-section measurements,
rather than weighted integrals, we discuss in details the experimental 
measurements to be performed and describe how to obtain qualitative 
information from the data. We use the QCD parton model language, at leading 
twist, allowing for transverse quark motion in the fragmentation process 
only (Collins effect \cite{col}); we do not discuss contributions from higher 
twist distribution or correlation functions.

\vskip 18pt
\nd
{\bf 2. Collins fragmentation function}
\vskip 6pt
Let us consider a quark with momentum $\bfp_q$ and a {\it transverse}  
polarization vector $\bfP_q$ ($\bfp_q \cdot \bfP_q = 0$) which fragments 
into a hadron with momentum $\bfp_h = z\bfp_q + \bfp_T$ 
($\bfp_q \cdot \bfp_T = 0$): according to Collins suggestion \cite{col}
the fragmentation function for the $q \to h + X$ process can be written as
\beq 
D_{h/q}(\bfp_q, \bfP_q; z, \bfp_T) = D_{h/q}(z, p_T) + \frac 12 \> 
\Delta^ND_{h/q}(z, p_T) \> \frac{\bfP_q \cdot (\bfp_q \times \bfp_T)}
{|\bfp_q \times \bfp_T|} \label{colfn}
\eeq
where $D_{h/q}(z, p_T)$ is the unpolarized fragmentation function.
Notice that -- as required by parity invariance -- the only component 
of the polarization vector which contributes to the spin dependent part
of $D$ is that perpendicular to the $q-h$ plane; in general one has,
see Fig. 1:
\beq
\bfP_q \cdot 
\frac{\bfp_q \times \bfp_T} {|\bfp_q \times \bfp_T|}
= P_q \sin(\varphi_{P_q} - \varphi_h) \equiv P_q \sin\Phi_C \>, \label{colan}
\eeq
where $P_q = |\bfP_q|$ and we have defined the {\it Collins angle}
$\Phi_C \equiv \varphi_{P_q} - \varphi_h$. It is obvious that any longitudinal 
(with respect to $\bfp_q$) component of $\bfP_q$ does not contribute
to Eq. (\ref{colfn}). 
   
When studying single spin asymmetries one considers differences of 
cross-sections with opposite transverse spins; by reversing the nucleon 
spin all polarization vectors, including those of quarks, change sign 
and the quantity which eventually contributes to single spin asymmetries is:
\beq 
D_{h/q}(\bfp_q, \bfP_q; z, \bfp_T) - D_{h/q}(\bfp_q, -\bfP_q; z, \bfp_T)
= \Delta^ND_{h/q}(z, p_T) \> \frac{\bfP_q \cdot (\bfp_q \times \bfp_T)}
{|\bfp_q \times \bfp_T|} \label{coldf}
\eeq
which implies the existence of a {\it quark analysing power}
for the fragmentation process $q \to h + X$:
\barr
A_q^h (\bfp_q, \bfP_q; z, \bfp_T) &=& \frac 
{D_{h/q}(\bfp_q, \bfP_q; z, \bfp_T) - D_{h/q}(\bfp_q, -\bfP_q; z, \bfp_T)}
{D_{h/q}(\bfp_q, \bfP_q; z, \bfp_T) + D_{h/q}(\bfp_q, -\bfP_q; z, \bfp_T)}
\label{aq} \\
&=&\frac {\Delta^ND_{h/q}(z, p_T)}   
{2\,D_{h/q}(z, p_T)} \> \frac {\bfP_q \cdot (\bfp_q \times \bfp_T)}
{|\bfp_q \times \bfp_T|} 
\equiv A_q^h (z, p_T) \> \frac {\bfP_q \cdot (\bfp_q \times \bfp_T)}
{|\bfp_q \times \bfp_T|} \> \cdot \nonumber 
\earr

A simple explicit expression of the function $\Delta^ND(z, 
\langle p_T \rangle)$ was obtained in Ref. \cite{noi2} by fitting the data 
on single spin asymmetries in $p \pup \to \pi X$ processes \cite{e704} under 
the assumption that they entirely originate from Collins effect alone.
It results that the quark analysing power $A_q^{\pi}(z, p_T)$ has to reach 
the limiting value $- 1$ at large $z$ in order to describe the data.

However, such a result depends also on the chosen expressions for the 
polarized quark distribution functions, in particular on their large $x$ 
behaviour. The same study was repeated in Ref. \cite{bl}, allowing for 
unusual large $x$ behaviours of the longitudinally polarized 
distribution functions $\Delta q(x)$ and imposing Soffer's inequality
bound \cite{sof} on the transversely polarized distribution functions 
$\Delta_Tq(x)$ or $h_1(x)$; the E704
data \cite{e704} could then be fitted leading 
to a constant, negative, $z$ independent, value of $A_q^{\pi}(z, p_T)$.

We discuss now how to extract information on $A_q^h(z,p_T)$ from semi-inclusive
DIS scattering processes, and use the first available data \cite{her,smc}
to obtain a lower bound estimate of its magnitude.

\vskip 18pt
\nd
{\bf 3. Single transverse spin asymmetries in \mbox{\boldmath
{$\gamma^* p^\uparrow \to \pi X$}}}
\vskip 6pt

Let us consider the semi-inclusive DIS process $\ell p \to \ell' \pi X$
in the $\gamma^* - p$ center of mass frame, with the kinematical 
configuration of Fig. 2, where we have chosen $x-z$ as the lepton
scattering plane with the $\gamma^*$ moving along the positive $z$-axis
and with:
\beq
\hat{\bfy} = \frac{\bfel \times \bfel'}{|\bfel \times \bfel'|}
%\equiv \hat{\bfN} 
\quad\quad\quad \hat{\bfx} = \hat{\bfy} \times \hat{\bfz} \>. \label{ref} 
\eeq
Let 
\beq
x = \frac{Q^2}{2p \cdot q}\>, \quad\quad y = \frac{Q^2}{sx} \>, 
\quad\quad s = (\ell + p)^2 \>, \quad\quad z = \frac{p \cdot p_h}{p \cdot q} 
\>, \label{var}
\eeq
be the usual DIS variables.

In the configuration of Fig. 2  -- neglecting intrinsic motion of quarks 
inside the initial proton -- the elementary interaction, 
$\gamma^* + q, \bfP_q \to q', \bfP_{q'}$, takes place along the $\hat z$-axis:
$\bfP_q$ is the initial quark transverse spin-polarization vector and the 
final quark transverse spin-polarization $\bfP_{q'}$ is fixed by the 
QED elementary dynamics and is given by:
\barr
(\bfP_{q'})_x &=& \frac {-2(1-y)}{1 + (1-y)^2} \> (\bfP_q)_x \label{pqx} \\ 
(\bfP_{q'})_y &=& \frac {2(1-y)}{1 + (1-y)^2} \> (\bfP_q)_y \>. \label{pqy}  
\earr
Notice that changing $\bfP_q$ into $-\bfP_q$ changes the sign of $\bfP_{q'}$.
The factor $|\bfP_{q'}|/|\bfP_{q}| = 2(1-y)/[1+(1-y)^2] \equiv \hat D_{NN}$ 
is the so-called depolarization factor; apart from this magnitude reduction,
$\bfP_q$ and $\bfP_{q'}$ only differ for the sign of the $x$-component which
corresponds to the situation described in Fig. 3.
 
Let us consider the general case of a nucleon with some transverse
spin-polari\-za\-tion vector 
\beq
\bfP = P\,(\cos\varphi_P, \> \sin\varphi_P, \> 0) \>. \label{vecp}
\eeq
If we denote by $P_\uparrow$ ($P_\downarrow$) the probability for the 
nucleon to have spin parallel (antiparallel) to $\bfP$ the degree of 
transverse polarization of the nucleon is given by:
\beq
P = |\bfP| = P_\uparrow - P_\downarrow \>. \label{pol}
\eeq
   
The measured cross-sections for a proton with transverse spin-polarization 
vector $\bfP$ can be written as
\barr
\frac{d\sigma^{\ell + p,\bfP \to \ell' + h + X}}
{dx \, dy \, dz \, d\bfp_T} &=& 
\sum_q \Biggl[ f_\uparrow^{\bfP}(x) \> \frac{d\hat\sigma^{\uparrow}_q}{dy} \>
D_{h/q}(\bfP_{q'}; z, \bfp_T)  \nonumber \\
&+& f_\downarrow^{\bfP}(x) \> \frac{d\hat\sigma^{\downarrow}_q}{dy} \>
D_{h/q}(-\bfP_{q'}; z, \bfp_T) \Biggr]
\label{crs+} 
\earr
where $f_{\uparrow(\downarrow)}^{\bfP}$ is the density number
of transversely polarized quarks with polarization $\uparrow \> = \hat{\bfP}$
($\downarrow \> = -\hat{\bfP}$) inside a proton with polarization vector 
$\bfP$ (notice the difference between $\bfP$ and $\hat{\bfP} = \bfP/P$:
\beq
f_\uparrow^{\bfP} = P_\uparrow \, f_\uparrow^\uparrow +  
P_\downarrow \, f_\uparrow^\downarrow \quad\quad
f_\downarrow^{\bfP} = P_\uparrow \, f_\downarrow^\uparrow +  
P_\downarrow \, f_\downarrow^\downarrow \>. \label{fp}
\eeq
$D_{h/q}(\pm \bfP_{q'}; z, \bfp_T)$ is the fragmentation function of the
final quark with a polarization vector $\pm \bfP_{q'}$ resulting from the 
$\ell + q,\pm\hat{\bfP} \to \ell' + q'$ interaction:
\beq
D_{h/q}(\bfP_{q'}; z, \bfp_T) = \frac
{d\hat\sigma_q^{\uparrow \to \uparrow} \> D_{h/\qup} +
 d\hat\sigma_q^{\uparrow \to \downarrow} \> D_{h/\qdown}}
{d\hat\sigma_q^{\uparrow \to \uparrow} + 
 d\hat\sigma_q^{\uparrow \to \downarrow}} \label{d+pqp}
\eeq
and
\beq
D_{h/q}(-\bfP_{q'}; z, \bfp_T) = \frac
{d\hat\sigma_q^{\downarrow \to \uparrow} \> D_{h/\qup} +
 d\hat\sigma_q^{\downarrow \to \downarrow} \> D_{h/\qdown}}
{d\hat\sigma_q^{\downarrow \to \uparrow} + 
 d\hat\sigma_q^{\downarrow \to \downarrow}} \> , \label{d-pqp}
\eeq
where $d\hat\sigma_q^{\uparrow \to \uparrow}$ stands for
$d\hat\sigma^{\ell + q,\uparrow \to \ell' + q,\uparrow}/dy$, and so on;
$\bfP_{q'}$ is explicitely given in Eqs. (\ref{pqx}) and (\ref{pqy}),
taking $\bfP_q = \hat{\bfP}$. 
As there is no single spin asymmetry in the elementary interaction one has 
$d\hat\sigma_q^{\uparrow \to \uparrow} + 
d\hat\sigma_q^{\uparrow \to \downarrow} =
d\hat\sigma_q^{\uparrow} = 
d\hat\sigma_q^{\downarrow \to \uparrow} + 
d\hat\sigma_q^{\downarrow \to \downarrow} =
d\hat\sigma_q^{\downarrow} = d\hat\sigma^{unp}_q$.
Notice also that $d\hat\sigma_q$ depends on the quark flavour only via its 
factorized charge, $d\hat\sigma_q = e_q^2 \, d\hat\sigma$.  

Similarly
\barr
\frac{d\sigma^{\ell + p,-\bfP \to \ell' + h + X}}
{dx \, dy \, dz \, d\bfp_T} &=& 
\sum_q \Biggl[ f_\uparrow^{-\bfP}(x) \> \frac{d\hat\sigma^{\uparrow}_q}{dy} \> 
D_{h/q}(\bfP_{q'}; z, \bfp_T) \nonumber \\
&+& f_\downarrow^{-\bfP}(x) \> \frac{d\hat\sigma^{\downarrow}_q}{dy} \>
D_{h/q}(-\bfP_{q'}; z, \bfp_T) \Biggr] \>.
\label{crs-} 
\earr

From Eqs. (\ref{coldf}), (\ref{crs+}) and (\ref{crs-})
one has:
\barr
2 \> \frac{d\sigma^{\ell + p \to \ell' + h + X}}
{dx \, dy \, dz \, d\bfp_T} &=& 
\frac{d\sigma^{\ell + p,\bfP \to \ell' + h + X}}
{dx \, dy \, dz \, d\bfp_T} + 
\frac{d\sigma^{\ell + p,-\bfP \to \ell' + h + X}}
{dx \, dy \, dz \, d\bfp_T} \nonumber \\ 
&=& \sum_q f_{q/p}(x) \> \frac{d\hat\sigma_q}{dy} \> 
2\,D_{h/q}(z, p_T) \label{crsun}
\earr 
and
\barr
&&\frac{d\sigma^{\ell + p,\bfP \to \ell' + h + X}}
{dx \, dy \, dz \, d\bfp_T} - 
\frac{d\sigma^{\ell + p,-\bfP \to \ell' + h + X}}
{dx \, dy \, dz \, d\bfp_T} \nonumber \\  
&=& \sum_q h_{1q}(x) \> \frac{d\hat\sigma_q}{dy} \> 
\Delta^ND_{h/q}(z, p_T) \> P \> \frac{\bfP_{q'} \cdot (\bfp_{q'} \times 
\bfp_T)} {|\bfp_{q'} \times \bfp_T|} \> \cdot \label{crsas}
\earr 
where $h_{1q} =  f_\uparrow^\uparrow - f_\uparrow^\downarrow$ is the
transverse spin distribution function. The single spin asymmetry for the 
process $\ell \pup \to \ell' h X$ is then given by:

\barr  
A^h_N(x,y,z,\bfP_{q'},\bfp_T) &=&
 \frac{d\sigma^{\ell + p,\bfP \to \ell' + h + X}
      -d\sigma^{\ell + p,-\bfP \to \ell' + h + X}}
      {d\sigma^{\ell + p,\bfP \to \ell' + h + X}
      +d\sigma^{\ell + p,-\bfP \to \ell' + h + X}} \nonumber \\
&=& \frac{\sum_q e_q^2 \, h_{1q}(x) \> \Delta^ND_{h/q}(z, p_T)}
{2\sum_q e_q^2 \, f_{q/p}(x) \> D_{h/q}(z, p_T)} \> P \>
\frac{\bfP_{q'} \cdot (\bfp_{q'} \times \bfp_T)}
{|\bfp_{q'} \times \bfp_T|} \> \cdot \label{asym1}
\earr

Using Eqs. (\ref{colan}) and (\ref{pqx}), (\ref{pqy}) the above asymmetry 
can be written as 
\beq
A^h_N(x,y,z,\Phi_C, p_T) =
\frac{\sum_q e_q^2 \, h_{1q}(x) \> \Delta^ND_{h/q}(z, p_T)}
{2\sum_q e_q^2 \, f_{q/p}(x) \> D_{h/q}(z, p_T)} \>
\frac{2(1-y)} {1 + (1-y)^2} \> P \> \sin\Phi_C \>. \label{asym2} 
\eeq

The value of $\sin\Phi_C$ depends on the direction of $\bfP_{q'}$:
from Eqs. (\ref{pqx}) and (\ref{pqy}) or their graphical 
representation in Fig. 3 one can express the Collins angle $\Phi_C$
in terms of measurable angles:
\beq
\Phi_C = \pi - \varphi_P - \varphi_h \>, \label{phic}
\eeq
where $\varphi_P$ and $\varphi_h$ are respectively the azimuthal angles
of the proton polarization vector and of the produced hadron; we 
always refer to the choice of axes of Fig. 2.

Some comments might be useful.
\begin{itemize}
\item
Eq. (\ref{asym2}) simplifies for pion production if one assumes that 
transverse sea quark polarization in a nucleon is negligible, 
$h_{1\bar q} \simeq 0$, and if one neglects fragmentation of non-valence 
quarks into a pion; that is, if one assumes, using also isospin and charge 
conjugation invariance:
\barr
D_{\pi^+/u} &=& D_{\pi^-/d} = D_{\pi^+/\bar d} = D_{\pi^-/\bar u} 
\equiv  D_{\pi/q} \label{dfav1} \\
2D_{\pi^0/u} &=& 2D_{\pi^0/d} = 2D_{\pi^0/\bar u} = 2D_{\pi^0/\bar d}
\equiv D_{\pi/q} \label{dfav2} \\
D_{\pi^+/d} &=& D_{\pi^-/u} = D_{\pi^+/\bar u} = D_{\pi^-/\bar d} \simeq 0
\label{dunf}
\earr
and similarly for the $\Delta^ND$. Eq. (\ref{dunf}) is justified at the 
large $z$ values for which data are available; moreover, for $\pi^+$,
the contribution of $d$ and $\bar u$ quarks are further suppressed
by the corresponding smaller distribution functions. 

In this case Eq. (\ref{asym2}) reads ($i = +, -, 0$):
\beq
A^{\pi^i}_N(x,y,z,\Phi_C, p_T) = 
\frac{h_i(x)}{f_i(x)} \> A_q^\pi(z, p_T) \> 
%\frac{\Delta^ND_{\pi/q}(z, p_T)}{2 \, D_{\pi/q}(z, p_T)} \>
\frac{2(1-y)} {1 + (1-y)^2} \> P \> \sin\Phi_C \label{aspi} 
\eeq
where:
\beq
i = + : \quad h_+ = 4h_{1u} \quad\quad 
f_+ = 4f_{u/p} +  f_{\bar d/p} \label{hf+} 
\eeq
\beq 
i = - : \quad h_- =  h_{1d} \quad\quad 
f_- =  f_{d/p} + 4f_{\bar u/p}  \label{hf-} 
\eeq
\beq 
i = 0 : \quad h_0 = 4h_{1u} + h_{1d} \quad\quad 
f_0 = 4f_{u/p} + f_{d/p} + 4f_{\bar u/p} + f_{\bar d/p} \>. \label{hf0}
\eeq
%
%\beq
%A^{\pi^-}_N(x,y,z,\Phi_C, p_T) =
%\frac{h_{1d}(x)}{f_{d/p}(x) + 4f_{\bar u/p}(x)} \> A_q^\pi(z, p_T) \> 
%\frac{\Delta^ND_{\pi/q}(z, p_T)}{2 \, D_{\pi/q}(z, p_T)} \>
%\frac{2(1-y)} {1 + (1-y)^2} \> P \> \sin\Phi_C \>. \label{aspi-} 
%\eeq
%
%
%\barr
%A^{\pi^0}_N(x,y,z,\Phi_C, p_T) =
%\frac{4h_{1d}(x) + h_{1d}(x)}
%{4f_{u/d}(x) + f_{d/p}(x) + 4f_{\bar u/p}(x) + f_{\bar d/p}(x)} 
%\> A_q^\pi(z, p_T) \> 
%\frac{\Delta^ND_{\pi/q}(z, p_T)}{2 \, D_{\pi/q}(z, p_T)} \>
%\frac{2(1-y)} {1 + (1-y)^2} \> P \> \sin\Phi_C \>. \label{aspi0} 
%\eeq
 
\item
The asymmetries (\ref{aspi}) are factorized into the
product of five different physical quantities, each of which is smaller in
magnitude than 1; it is remarkable that the existing data \cite{her, smc}, 
hint at values of $A_N$ of the order of a few percents. Out of 
the five factors, three -- the transverse polarization $P$, the depolarization
factor and $\sin\Phi_C$ -- are known or measurable separately; the other 
two -- $h_i/f_i$ and the quark analysing power 
$A_q^\pi = \Delta^ND_{\pi/q}/2D_{\pi/q}$ -- are 
not known. The measured values of $A_N$ yield information on the {\it product}
of these two quantities. However -- and we shall attempt this in the sequel --
some upper bounds for $h_1$ are known \cite{sof}, and we should be able to 
obtain lower bound estimates for the size of the quark analysing power.

\item
In order to cumulate data collected at different kinematical values, 
one should integrate the cross-sections (\ref{crs+}) and (\ref{crs-}),
for example over the relevant $x$ and $y$ regions; in this case one obtains:
\barr
&&A^h_N(z, \Phi_C, p_T) =
 \frac{\int dx \, dy \> [d\sigma^{\ell + p,\bfP \to \ell' + h + X}
      -d\sigma^{\ell + p,-\bfP \to \ell' + h + X}]}
      {\int dx \, dy \> [d\sigma^{\ell + p,\bfP \to \ell' + h + X}
      +d\sigma^{\ell + p,-\bfP \to \ell' + h + X}]} \nonumber \\
&=& 
\frac{\sum_q \int dx \, dy \, e_q^2 \, h_{1q}(x) \> 2(1-y)/(xy^2) \> 
\Delta^ND_{h/q}(z, p_T) \> P \> \sin\Phi_C}
{2\sum_q \int dx \, dy \, e_q^2 \, f_{q/p}(x) \> (1 + (1-y)^2)/(xy^2) \>
D_{h/q}(z, p_T)} \> \cdot \label{asint}  
\earr
Notice that $P$ depends on $x$ and $y$, as we shall
discuss in the next Section. 

The analogue of Eq. (\ref{aspi}) can be written as:
\beq
A^{\pi^i}_N(z, \Phi_C, p_T) = \frac{I_{N_i}}{I_{D_i}} \> A_q^\pi(z, p_T) \>
%\frac{\Delta^ND_{\pi/q}(z, p_T)}{2D_{\pi/q}(z, p_T)} \> 
\sin\Phi_C \label{aspii}
\eeq
with 
\beq
I_{N_i} = \int dx \, dy \, h_{i}(x) \> \frac{2(1-y)}{xy^2} \> P \label{ini}
%\left[ 1 - \frac{2m_N^2 \, x(1-y)}{sy} \right] \> \label{in}
\eeq
and
\beq
I_{D_i} = \sum_q \int dx \, dy \, f_{i}(x) \> \frac{1 + (1-y)^2}{xy^2} \>, 
\label{id}  
\eeq
where $h_i$ and $f_i$ are given in Eqs. (\ref{hf+})-(\ref{hf0}). 
In factorizing the quark analysing power $A_q^\pi$ in Eq. (\ref{aspii}) 
we have neglected the smooth $y$ dependence of the fragmentation functions 
induced by QCD $Q^2$-evolution. 
 
\end{itemize}

\vskip 18pt
\nd
{\bf 4. Measurements of $\bfAN$ in particular cases}
\vskip 6pt

Eq. (\ref{asym2}) holds at leading twist  -- contributions decreasing as
inverse powers of $Q$ have been neglected -- and leading order in the 
parton transverse motion, in the sense that we have taken into 
account quark transverse momenta only in places where neglecting them 
would give zero value for $A_N$: the quarks inside the initial fast proton 
are all collinear and the observed $\bfp_T$ of the pion arises entirely 
as transverse momentum in the fragmentation process.

In Eqs. (\ref{asym2}) and (\ref{asint}) $\bfP$ is the proton polarization 
transverse to its direction in the $\gamma^*-p$ center of mass frame 
($\bfP \cdot \bfq = 0$): it is the only component of the polarization which 
contributes to $A_N$. In realistic situations the proton spin-polarization 
vector is not perpendicular to $\bfq$; there are two typical configurations 
in performed or planned experiments:

\begin{enumerate}
\item[a)]     
The proton spin-polarization vector, $\bfP_L$, is {\it longitudinal} with
respect to the lepton direction in the laboratory frame, where the proton
is at rest (Fig. 4).

\item[b)]
The proton spin-polarization vector, $\bfP_T$, is {\it transverse} with
respect to the lepton direction in the laboratory frame, where the proton
is at rest (Fig. 5).
\end{enumerate}

Let us consider these two cases separately.
\vskip 6pt
\nd
a) - {\it Longitudinal proton polarization}, $\bfP_L$
\vskip 6pt

In this case the magnitude of the active component of the proton 
spin-polari\-za\-tion vector -- {\it i.e.} the component which contributes 
to $A_N$ -- is given by (see Fig. 4):
\beq
P = P_L \> \sin\theta_\gamma \>, \label{ppl}
\eeq
with ($m_N$ is the proton mass)
\beq
\sin\theta_\gamma = 2m_N \> \left[ \frac{
(s-m_N^2) \, xy(1-y) - m_N^2 \, x^2y^2}
{(s-m_N^2)^2 \, y^2 + 4m_N^2 \, (s-m_N^2) \, xy} \right]^{1/2}
\simeq \frac{2m_N \, x}{Q} \> \sqrt{1-y} \>. \label{stg}
\eeq
In the last expression we have neglected terms of $O(m_N^3/Q^3)$.

Eq. (\ref{stg}) shows how a longitudinal nucleon polarization
can contribute to single transverse spin asymmetries in 
semi-inclusive DIS processes only at moderate values of $Q$ and in events 
for which the factor $2 \, x \sqrt{1-y}$ is not too small; 
this is the case of HERMES measurements \cite{her}. It also indicates that
higher twist contributions might be as important as leading twist ones,
as discussed in Refs. \cite{dan, oga}.   

Fig. 4 shows how the azimuthal angle of $\bfP_L$ is 0, so that, 
from Eq. (\ref{phic}),
\beq
\Phi_C = \pi - \varphi_h \>. \label{phica}
\eeq

The factor $P \> \sin\Phi_C$ for a longitudinal proton polarization,
assuming $P_L = 1$, is then 
\beq
P \> \sin\Phi_C \simeq \frac {2m_N \, x}{Q} \> \sqrt{1-y} \>
\sin\varphi_h \>. \label{psca}
\eeq
\vskip 6pt
\nd
b) - {\it Transverse proton polarization}, $\bfP_T$
\vskip 6pt

In this case it is more elaborate to obtain the value of $P$, the magnitude
of the polarization vector component orthogonal to $\bfq$, starting from 
a polarization vector $\bfP_T$ orthogonal to the lepton momentum, in 
the laboratory frame (see Fig. 5); one can show that
\beq
P = P_T \> \biggl[ \cos^2\theta_\gamma \cos^2\phi_P^{\ell'} + 
\sin^2\phi_P^{\ell'} \biggr]^{1/2} \label{ppt}
\eeq
where $\phi_P^{\ell'}$ is the difference between the azimuthal angle of 
$\bfP_T$  and $\bfel'$, measured around the initial lepton direction,
in the frame in which the proton is at rest, and 
\beq
\cos\theta_\gamma = \frac{(s-m_N^2) \, y + 2m_N^2 \, xy}
{[(s-m_N^2)^2 \, y^2 + 4m_N^2 \, (s-m_N^2) \, xy]^{1/2}}
\simeq 1 - \frac {2m_N^2 \, x^2}{Q^2} \> (1-y) \>. \label{ctg}
\eeq
In the last expression we have neglected terms of $O(m_N^4/Q^4)$.

Eq. (\ref{ppt}) shows that, in this case, $P$ varies with $\phi_P^{\ell'}$, 
that is with the orientation of the initial transverse polarization
around $\bfel$, which is known experimentally, event by event. However,
one sees also that, with varying $\phi_P^{\ell'}$, one always has 
$P_T \> \cos\theta_\gamma \leq P \leq P_T$ and that, on the average
(assuming all values of $\phi_P^{\ell'}$ to be equally probable):
\beq
\langle P^2 \rangle = \frac{P_T^2}{2} \> (1 + \cos^2\theta_\gamma) \>. 
\label{avp}
\eeq

The factor $P \> \sin\Phi_C$ for a transverse proton polarization,
assuming $P_T = 1$, can then be approximated, using Eqs. (\ref{phic}) and 
(\ref{avp}), as
\beq
P \> \sin\Phi_C \simeq \left[ 1 - \frac {2m_N^2 \, x^2}{Q^2} \> (1-y) 
\right] \> \sin(\varphi_P + \varphi_h) \>. \label{pscb}
\eeq
Notice anyway from Eq. (\ref{ctg}) that $\cos\theta_\gamma$, for a $Q^2$ of
a few (GeV/$c)^2$, is very close to 1 so that in practice $P \simeq P_T$.

By comparing Eqs. (\ref{psca}) and (\ref{pscb}) one sees that for 
longitudinally polarized protons even the leading twist contribution
to $A_N$, Eq. (\ref{asym2}), is suppressed by powers of $m_N/Q$, so that
higher twist contributions might be equally
important \cite{dan, oga}. For transversely polarized protons, instead,
the leading twist contribution is dominant, as in this case $P \simeq P_T$. 

\vskip 18pt
\nd
{\bf 5. Estimates of the quark analysing power}
\vskip 6pt

We consider now the case of $\pi^+$ and $\pi^-$ production with a transversely
polarized proton, as measured by SMC collaboration \cite{smc}, or 
with a longitudinally polarized proton, as measured by HERMES collaboration
\cite{her}.

The leading twist expressions for $A_N^{\pi^i}$, Eq. (\ref{aspi}) or 
its integrated versions, Eq. (\ref{aspii}), depend on 
the transverse spin distribution functions $h_{1q}$ which are not known:
however, one can derive an upper limit for the magnitude of $h_i$
or $I_{N_i}$ by exploiting the accurate knowledge of the 
unpolarized distribution functions, $q = f_{q/p}$, of the longitudinally
polarized ones, $\Delta q$, and the Soffer bound on $h_{1q}$ \cite{sof}:
\beq
|h_{1q}| \leq \frac12 \> (q + \Delta q) \>. \label{sofb}
\eeq

This allows, by comparison with the available data on $A_N^\pi$,
to derive a {\it lower bound estimate} for the quark analysing power 
$A_q^\pi$. We do this, considering separately the SMC 
and HERMES data, and making some further comments on the latter.

\vskip 6pt
\nd
{\it SMC data, transverse proton polarization, Fig. 5}
\vskip 6pt

Ref. \cite{smc} reports preliminary results on $A_N^h$ for positive
and negative hadrons (mainly pions) produced in the DIS scattering 
of unpolarized muons off a transversely polarized proton target at:
\beq
\langle x \rangle \simeq 0.08 \quad\quad
\langle Q^2 \rangle \simeq 5 \> (\mbox{GeV}/c)^2 \quad\quad  
\langle z \rangle \simeq 0.45 \>, \label{kin1}
\eeq 
which implies $\langle y \rangle \simeq 0.33$, as $s = 188.5$ (GeV$)^2$.
They have selected two data sets with $\langle p_T \rangle = 0.5$ GeV/$c$
and $\langle p_T \rangle = 0.8$ GeV/$c$; it is interesting to note 
that $|A_N^h|$ increases (moderately) with $p_T$. They present results
for the asymmetry divided by the depolarization factor and for the total 
amount of events they have: 
\barr
\left( A_N^{\pi^+} \right)_H = 
\frac{1 + (1-y)^2}{2(1-y)} \> A_N^{\pi^+} &=& - \> (0.11 \pm 0.06) \>
\sin(\varphi_P + \varphi_h) \label{smc+} \\ 
\left( A_N^{\pi^-} \right)_H = 
\frac{1 + (1-y)^2}{2(1-y)} \> A_N^{\pi^-} &=& + \> (0.02 \pm 0.06) \>
\sin(\varphi_P + \varphi_h) \>. \label{smc-} 
\earr
Notice that the choice of axes in Ref. \cite{smc} is the same as ours,
but they define the Collins angle [see their Eq. (2)] with an {\it opposite} 
sign with respect to our definition, Eqs. (\ref{colan}) and (\ref{phic}).

Evaluating the depolarization factor at $\langle y \rangle \simeq 0.33$,
Eqs. (\ref{smc+}) and (\ref{smc-}) imply:      
\barr
A_N^{\pi^+} &=& - \> (0.10 \pm 0.06) \> \sin(\varphi_P + \varphi_h) 
\label{smc+1} \\
A_N^{\pi^-} &=& + \> (0.02 \pm 0.06) \> \sin(\varphi_P + \varphi_h) \>. 
\label{smc-1} 
\earr

On the other hand Eqs. (\ref{aspi})-(\ref{hf-}), (\ref{pscb}) and 
(\ref{sofb}), evaluated at the average kinematical values of Eq. (\ref{kin1}), 
with the polarized and unpolarized quark distributions of Ref. \cite{grsv}, 
yield:
\barr
|A_N^{\pi^+}(\langle z \rangle, \Phi_C, \langle p_T \rangle)| &\lsim& 0.41 \> 
|A_q^\pi(\langle z \rangle, \langle p_T \rangle) \>
%\frac{|\Delta^ND_{\pi/q}(\langle z \rangle, \langle p_T \rangle)|}
%{2D_{\pi/q}(\langle z \rangle, \langle p_T \rangle)} \> 
\sin(\varphi_P + \varphi_h)| 
\label{smc+2} \\
|A_N^{\pi^-}(\langle z \rangle, \Phi_C, \langle p_T \rangle)| &\lsim& 0.11 \> 
|A_q^\pi(\langle z \rangle, \langle p_T \rangle) \>
%\frac{|\Delta^ND_{\pi/q}(\langle z \rangle, \langle p_T \rangle)|}
%{2D_{\pi/q}(\langle z \rangle, \langle p_T \rangle)} \> 
\sin(\varphi_P + \varphi_h)| \>.
\label{smc-2} 
\earr

From Eqs. (\ref{smc+2}) and (\ref{smc+1}) one has, assuming $h_{1u}$ to
be positive, as all models indicate:
\beq
%\frac{\Delta^ND_{\pi/q}(\langle z \rangle, \langle p_T \rangle)}
%{2D_{\pi/q}(\langle z \rangle, \langle p_T \rangle)} 
A_q^\pi(\langle z \rangle, \langle p_T \rangle) \lsim - (0.24 \pm 0.15)
\quad\quad \langle z \rangle \simeq 0.45 \>, \quad 
\langle p_T \rangle \simeq 0.65 \> \mbox{GeV}/c \>.
\label{res1}
\eeq
The experimental result (\ref{smc-1}) is not statistically significant
to allow an independent evaluation of a bound on $A_q^\pi$;
%$\Delta^ND_{\pi/q}/2D_{\pi/q}$; 
however Eqs. (\ref{smc-2}) and (\ref{res1}) 
are in agreement with (\ref{smc-1}).     

The estimate (\ref{res1}) is obtained by taking average values of the 
various kinematical variables, according to the indications of 
Ref. \cite{her}; we have also redone the above evaluation by explicitely 
performing the $x$ and $y$ integration,
\barr
&&\left( A^{\pi^i}_N(z, \Phi_C, p_T) \right)_H =
 \frac{\int dx \, dy \> [d\sigma^{\ell + p,\bfP \to \ell' + \pi^i + X}
      -d\sigma^{\ell + p,-\bfP \to \ell' + \pi^i + X}] \> (\hat D_{NN})^{-1}}
      {\int dx \, dy \> [d\sigma^{\ell + p,\bfP \to \ell' + h + X}
      +d\sigma^{\ell + p,-\bfP \to \ell' + h + X}]} \nonumber \\
&=& 
\frac{\int dx \, dy \, h_{i}(x) \> [1 + (1-y)^2]/(xy^2) \> P}
{\int dx \, dy \, f_{i}(x) \> [1 + (1-y)^2]/(xy^2)} \> 
\> A_q^{\pi}(z, p_T) \> \sin\Phi_C \>, \label{asinth}  
\earr
over the range $x \geq 0.0053/y = Q^2_{min}/(ys), \quad 0.1 \leq y \leq 0.7, 
\quad Q^2 \geq 1$ (GeV/$c)^2$ (results are almost independent of the upper 
$x$ limit). 

From Eqs. (\ref{asinth}), (\ref{hf+}), (\ref{hf-}), (\ref{pscb}) and the
bound (\ref{sofb}), we obtain now: 
\barr
\left| \left( A_N^{\pi^+}(\langle z \rangle, \Phi_C, \langle p_T \rangle)
\right)_H \right| 
&\lsim& 0.42 \> |A_q^\pi(\langle z \rangle, \langle p_T \rangle) \>
%\frac{|\Delta^ND_{\pi/q}(\langle z \rangle, \langle p_T \rangle)|}
%{2D_{\pi/q}(\langle z \rangle, \langle p_T \rangle)} \> 
\sin(\varphi_P + \varphi_h)| 
\label{smc+3} \\
\left| \left( A_N^{\pi^-}(\langle z \rangle, \Phi_C, \langle p_T \rangle)
\right)_H \right| 
&\lsim& 0.11 \> |A_q^\pi(\langle z \rangle, \langle p_T \rangle) \>
%\frac{|\Delta^ND_{\pi/q}(\langle z \rangle, \langle p_T \rangle)|}
%{2D_{\pi/q}(\langle z \rangle, \langle p_T \rangle)} \> 
\sin(\varphi_P + \varphi_h)| \>.
\label{smc-3} 
\earr

to be compared with Eqs. (\ref{smc+}) and (\ref{smc-}).
This yields,
\beq
%\frac{\Delta^ND_{\pi/q}(\langle z \rangle, \langle p_T \rangle)}
%{2D_{\pi/q}(\langle z \rangle, \langle p_T \rangle)} 
A_q^\pi(\langle z \rangle, \langle p_T \rangle) \lsim - (0.26 \pm 0.14)
\quad\quad \langle z \rangle \simeq 0.45 \>, \quad 
\langle p_T \rangle \simeq 0.65 \> \mbox{GeV}/c \>.
\label{res2}
\eeq
in agreement with (\ref{res1}). 

The conclusion induced by SMC data, Eq. (\ref{res1}) or (\ref{res2}), is 
remarkable and very interesting; the quark analysing power as defined in 
Eq. (\ref{aq}) is large (in magnitude) and negative; what is even more 
remarkable is the fact that the same qualitative conclusion was obtained in 
Ref. \cite{noi2}, by fitting the data on $p \pup \to \pi X$, assuming that 
the observed single spin asymmetries \cite{e704} are entirely due to
Collins effect. Actually, predictions for $A_N^\pi$ in DIS, using the
evaluation of $\Delta^ND_{\pi/q}$ of Ref. \cite{noi2} were computed in 
Ref. \cite{noi3}; however, they were given in different kinematical 
regions and cannot be compared directly with the data now available.

\vskip 6pt
\nd
{\it HERMES data, longitudinal proton polarization, Fig. 4}
\vskip 6pt
Ref. \cite{her} reports results on $A_N^\pi$ for positive and negative 
pions produced in the DIS scattering of unpolarized positrons off a 
transversely polarized proton target at $s = 52.6$ (GeV$)^2$ and in the 
kinematical ranges:
\beq
0.023 \leq x \leq 0.4 \quad\quad
0.1 \leq y \leq 0.85 \quad\quad Q^2 \geq \> 1\> (\mbox{GeV}/c)^2 
\quad\quad 0.2 \leq z \leq 0.7 \>. \label{kin3}
\eeq 

They give:
\barr
A_N^{\pi^+} &\simeq& + \> [0.022 \pm 0.004 (stat.) \pm 0.004(syst.)] 
\> \sin \varphi_h 
\label{her+1} \\
A_N^{\pi^-} &\simeq& - \> [0.001 \pm 0.005 (stat.) \pm 0.004(syst.)]  
\> \sin \varphi_h \>. 
\label{her-1} 
\earr
Although they do not take into account the change of direction of 
$\bfP_q$ during the elementary interaction, Eq. (\ref{pqx}),
their definition of $\sin\Phi_C$ agrees with ours, Eq. (\ref{phica}).
 
There seems to be a sign discrepancy between the SMC and HERMES data;
this should be investigated carefully and a consistent definition of
the Collins angle used in all experiments. 

One should not forget the previous comments about the fact that other 
higher twist contributions, like those related to the quark transverse 
spin distribution inside a longitudinally polarized proton
\cite{kot, dan, bm}, are expected to be as important as 
the leading twist ones, due to Eq. ({\ref{stg}): this might indeed 
change the sign of $A_N$ observed by HERMES, with respect to that observed 
by SMC, as they may originate from different contributions. A confirmation
of the sign difference would indeed confirm such situation.

Despite these considerations we proceed now as in the case of transversely 
polarized protons, and use the upper bound (\ref{sofb}), to obtain an upper
limit on the value of the leading twist Eq. (\ref{aspii}). This in turns 
results in a lower limit on $A_q^\pi$. 
Although this procedure is now much less reliable than in the case of
transversely polarized protons -- for which leading twist contributions
dominate -- still it should give us an order of magnitude 
estimate of the value of $|A_q^\pi|$, assuming that 
leading twist contributions are of the same order as non leading ones.  

From Eqs. (\ref{aspii})-(\ref{id}),(\ref{hf+})-(\ref{hf-}) and (\ref{psca}), 
together with the bound (\ref{sofb}), using the distribution functions of 
Ref. \cite{grsv}, and integrating over the kinematical range of 
Eq. (\ref{kin3}) one obtains:
\barr
|A_N^{\pi^+}(z, \Phi_C, p_T)| &\lsim& 0.111 \> |A_q^\pi(z, p_T) \> 
%\frac{|\Delta^ND_{\pi/q}(z, p_T)|} {2D_{\pi/q}(z, p_T)} \> 
\sin\varphi_h| \label{her+2} \\
|A_N^{\pi^-}(z, \Phi_C, p_T)| &\lsim& 0.026 \> |A_q^\pi(z, p_T) \>
%\frac{|\Delta^ND_{\pi/q}(z, p_T)|} {2D_{\pi/q}(z, p_T)} \> 
\sin\varphi_h| \>. \label{her-2}
\earr
By comparing Eqs. (\ref{her+2}) and (\ref{her+1}) one has:
\beq
%\frac{|\Delta^ND_{\pi/q}(z, p_T)|} {2D_{\pi/q}(z, p_T)} 
|A_q^\pi(z, p_T)| \gsim  0.20 \pm 0.04(stat.) \pm 0.04(syst.) 
\quad\quad\quad  z \geq 0.2 \>, \label{res3}
\eeq
which, with all uncertainties discussed, indicates again a large value of
$|A_q^\pi|$. Data on $\pi^-$ have much too large errors
to allow any significant numerical evaluation; however, Eqs. (\ref{res3}) and
(\ref{her-2}) are consistent with Eq. (\ref{her-1}).

\vskip 18pt
\nd
{\bf 6. Conclusions}
\vskip 6pt

The data \cite{her, smc} on the azimuthal dependence of pions produced in 
semi-inclusive DIS, with unpolarized leptons and polarized protons, are 
extremely interesting and, if confirmed, would reveal a new fundamental spin 
property of the quark fragmentation; this new property might also be 
responsible for other transverse single spin asymmetries observed in 
$pp$ inclusive processes \cite{e704}. A phenomenological approach \cite{noi2}
to these intriguing and subtle experimental measurements seems now indeed 
possible.

We have analysed SMC and HERMES data within the leading twist 
QCD-factorization scheme, allowing for intrinsic $\bfp_T$ and spin 
dependence in the fragmentation process of a transversely polarized
quark, the so-called Collins effect \cite{col}. This is certainly 
and largely the dominant contribution in case of transversely polarized
protons (SMC), whereas in case on longitudinal polarization (HERMES)
higher twist contributions might be equally or even more important. 

At leading twist, the measured single spin asymmetries depend on the 
product of two unknown functions, which contain the quark analysing power 
$A_q^\pi$, Eq. (\ref{aq}), and the transverse spin distributions $h_{1q}$. 
The first unavoidable conclusion from the data is that such a product has to 
be large. By exploiting the existence of an upper bound on $|h_{1q}|$ we are
able to derived a lower limit for the magnitude of $A_q^\pi$. 
The SMC data -- for which a leading twist analysis is well accurate -- 
indicate a large lower bound on $|A_q^\pi|$ and a negative sign for it,
in exact agreement with the findings of Ref. \cite{noi2}, where Collins 
effect was used to describe single transverse spin asymmetries in 
$p\pup \to \pi X$ processes. 

The HERMES data should be analysed taking into account higher twist 
contributions, as other authors do \cite{kot}-\cite{bm}; nevertheless,
they also point towards a large value of $|A_q^\pi|$. It is interesting
to notice that the asymmetries measured by SMC and HERMES, if a consistent
definition of the Collins angle is used, appear to differ in sign: the 
reported data have the same signs, but the definitions of $\sin\Phi_C$ used
by the two collaborations seem to be opposite. 
If such a difference persists one has to conclude that the origin of
the asymmetry measured by HERMES is still dominated by higher twist
contributions, and these give opposite (and larger) values with respect 
to the leading twist contributions.  

New data on single transverse spin asymmetries in semi-inclusive DIS and
other processes will be available from several experiments in progress 
or planned; a consistent way of understanding and predicting them in
terms of new, fundamental, spin and transverse momentum dependent, 
fragmentation and distribution functions appears possible.

\vskip 18pt 
\nd  
{\bf Acknowledgements}
\vskip 6pt
We would like to thank M. Boglione, A. Bravar and E. Leader for useful 
discussions.

\clearpage 

\begin{figure}[c]
\begin{center}
 \epsfig{figure=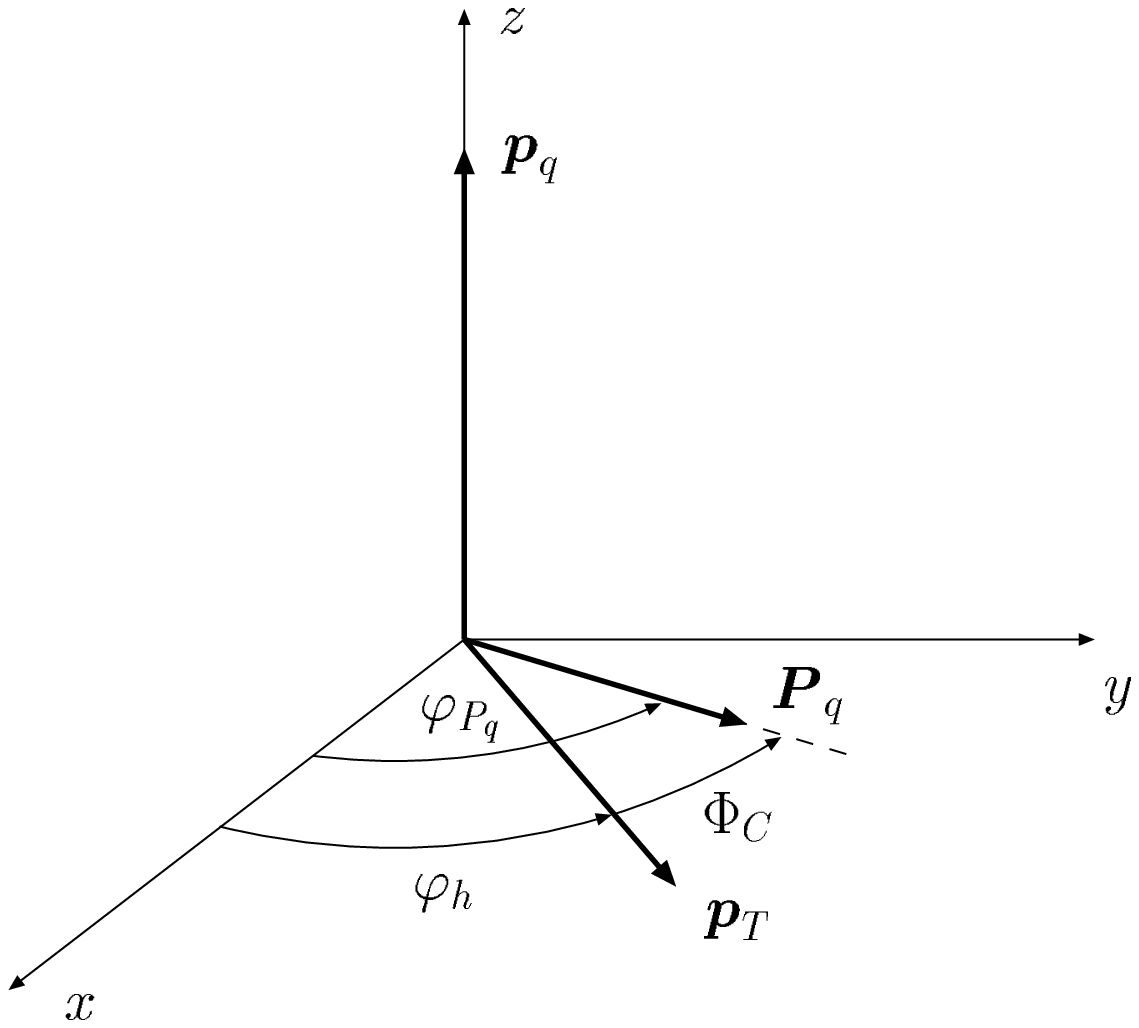,bbllx=150pt,bblly=315pt,bburx=500pt,%
bbury=665pt,width=14.0truecm,height=14.0truecm}
\end{center}
\begin{center}
 \begin{minipage}[c]{13cm}
 {\small {\bf Fig. 1:}
Definition of the Collins angle for the fragmentation of a
quark with momentum $\bfp_q$ and transverse spin-polarization vector
$\bfP_q$ into a hadron with momentum $\bfp_h = z\bfp_q + \bfp_T$:
$\bfP_q \cdot (\hat{\bfp}_q \times \hat{\bfp}_T) = P_q \> \sin(\varphi_{P_q}
- \varphi_h) \equiv P_q \> \sin\Phi_C$. }
 \end{minipage}
\end{center}
\end{figure}

\clearpage

\begin{figure}[c]
\begin{center}
 \epsfig{figure=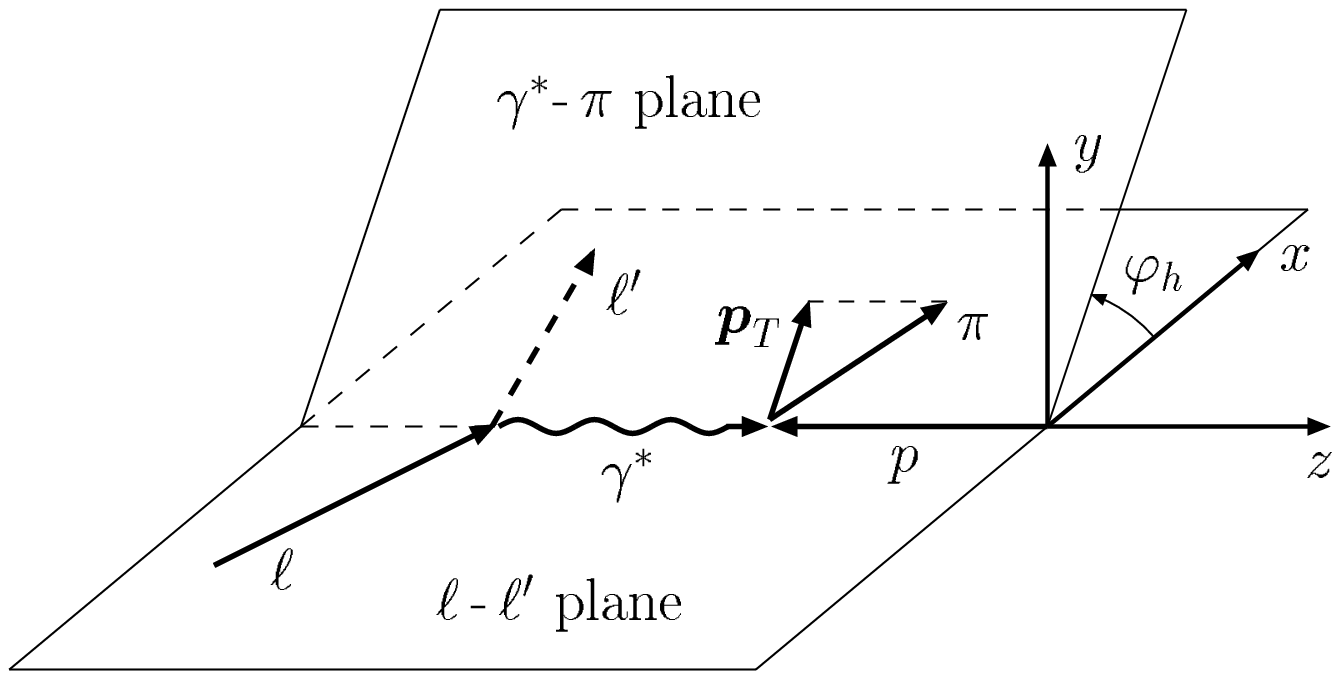,bbllx=80pt,bblly=360pt,bburx=500pt,%
bbury=600pt,width=14.0truecm,height=8.0truecm}
\end{center}
\begin{center}
 \begin{minipage}[c]{13cm}
 {\small {\bf Fig. 2:}
Definition of our kinematical configuration and choice of
the reference frame. }
 \end{minipage}
\end{center}
\end{figure}

\clearpage

\begin{figure}[c]
\begin{center}
 \epsfig{figure=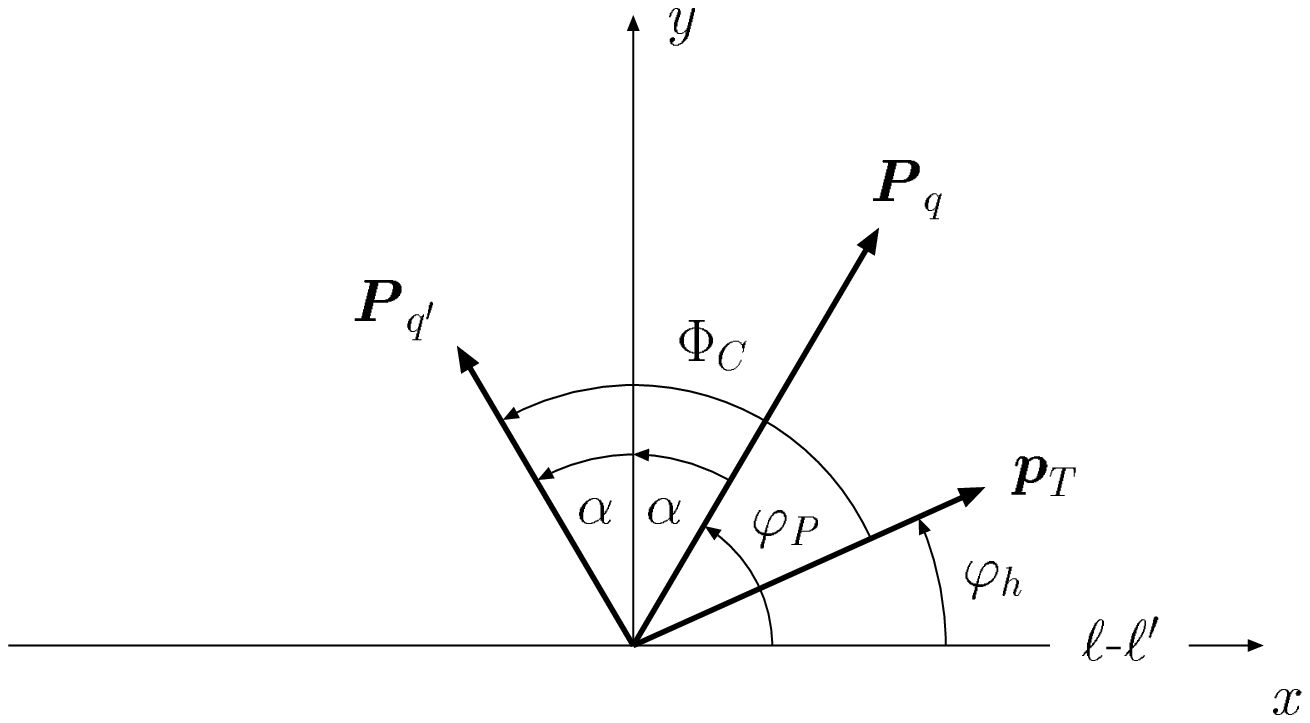,bbllx=100pt,bblly=400pt,bburx=500pt,%
bbury=660pt,width=14.0truecm,height=9.1truecm}
\end{center}
\begin{center}
 \begin{minipage}[c]{13cm}
 {\small {\bf Fig. 3:}
Transverse polarization vector $\bfP_{q'}$ resulting from the 
$\ell + q, \bfP_q \to \ell' + q'$ interaction: the final polarization
$\bfP_{q'}$ is reduced in magnitude by the depolarization factor, 
$|\bfP_{q'}| = \hat D_{NN} \> |\bfP_q|$ and is symmetric, with respect
to the $y$-axis, to $\bfP_q$. The $\gamma^*$ moves along the positive 
$z$-axis and the $\ell-\ell'$ plane is the $x-z$ plane. 
Notice that $\Phi_C = \varphi_{P_{q'}} - \varphi_h = \varphi_P + 2\alpha
- \varphi_h = \pi - \varphi_P - \varphi_h$.  }
 \end{minipage}
\end{center}
\end{figure}

\clearpage

\begin{figure}[c]
\begin{center}
 \epsfig{figure=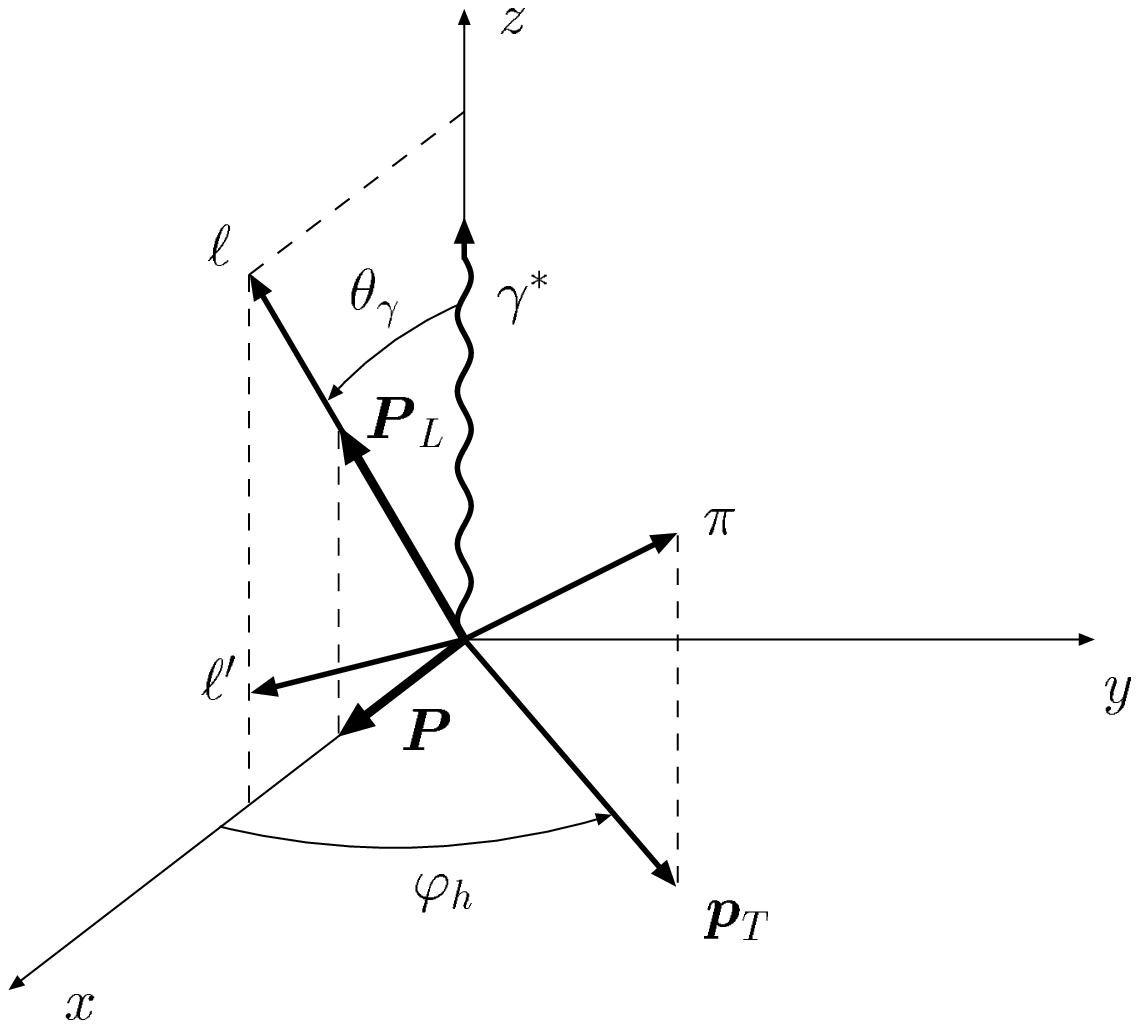,bbllx=150pt,bblly=315pt,bburx=500pt,%
bbury=665pt,width=14.0truecm,height=14.0truecm}
\end{center}
\begin{center}
 \begin{minipage}[c]{13cm}
 {\small {\bf Fig. 4:}
Kinematical configuration for HERMES experiment, in a frame
where the proton is at rest and $\gamma^*$ moves along the positive $z$-axis.
In this case $P = P_L \> \sin\theta_\gamma$ and the Collins angle simplifies 
to  $\Phi_C = \pi - \varphi_P - \varphi_h = \pi - \varphi_h$ (the polarization
vector of the final fragmenting quark points opposite to $\bfP$). }
 \end{minipage}
\end{center}
\end{figure}

\clearpage

\begin{figure}[c]
\begin{center}
 \epsfig{figure=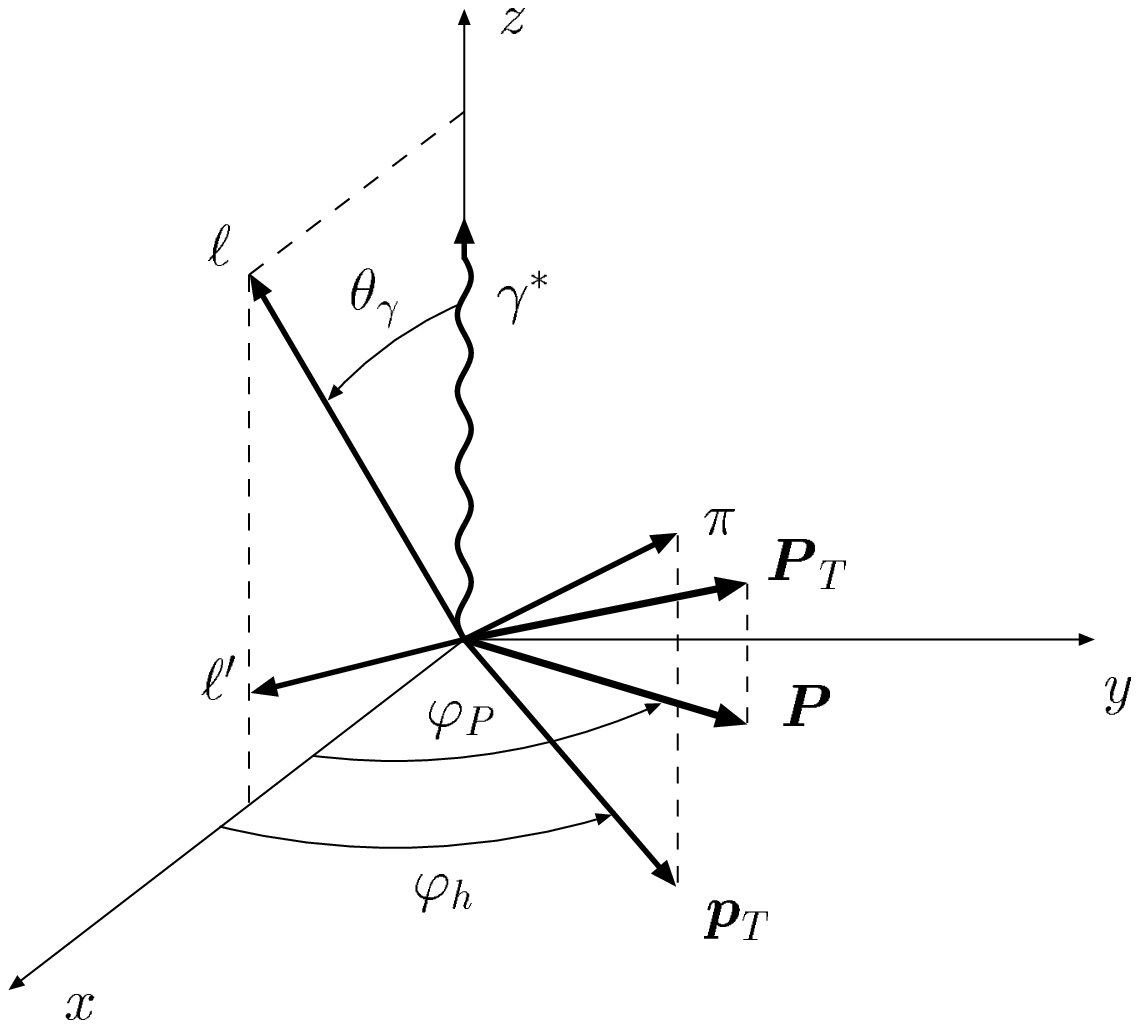,bbllx=150pt,bblly=315pt,bburx=500pt,%
bbury=665pt,width=14.0truecm,height=14.0truecm}
\end{center}
\begin{center}
 \begin{minipage}[c]{13cm}
 {\small {\bf Fig. 5:}
Kinematical configuration for SMC experiment, in a frame where 
the proton is at rest and $\gamma^*$ moves along the positive $z$-axis.
The proton polarization vector $\bfP_T$ is now orthogonal to $\bfel$; the 
polarization vector of the final fragmenting quark
has direction symmetric to $\bfP$ 
with respect to the $y$-axis and $\Phi_C = \pi - \varphi_P - \varphi_h$. }
 \end{minipage}
\end{center}
\end{figure}
\end{document}